# How the emotion's type and intensity affect rumor spreading

Yanli Li[a,b,*], Jing Ma[a], Fanshu Fang[a]

[a] *College of Economics and Management, Nanjing University of Aeronautics and Astronautics, Nanjing, Jiangsu, 211106, China.*

[b] *School of Cultural Management, Nanjing College of Communication, Nanjing, Jiangsu, 211172, China.*

HIGHLIGHTS

The type and intensity of emotions have important effects on DMs' behavior in rumor spreading.

Two theories of RDEU and evolutionary game are used.

Some significant emotional thresholds are found.

The evolutionary game results of strategies are summarised as five types.

**ABSTRACT:** The implication and contagion effect of emotion cannot be ignored in rumor spreading. This paper sheds light on how DMs' emotional type and intensity affect rumor spreading. Based on the theory of RDEU and evolutionary game, we construct an evolutionary game model of rumor spreading by considering emotions, which takes netizens and the government as the core subjects. Through MATLAB to simulate and reveal the influencing mechanism of DMs' emotional type and intensity on rumor spreading. The results indicate that the DMs' choice of strategy is not only affected by their own emotional preference and intensity but also affected by that of the other player in rumor spreading. Pessimism has a more significant influence than optimism on the stability of the evolutionary game, the government's emotion is more sensitive than netizens' to the game results, and the emotional intensity is proportional to the evolution speed. More importantly, some significant emotional thresholds are found, which can be used to predict the behavior of netizens, helping the government gain critical time to deal with rumors and avoid the Tacitus Trap crisis. Furthermore, the simulation results are summarized as five types: risk, opportunity, ideal, security, and opposition. We hope that this work is beneficial to the government's public governance.

## 1.Introduction

Just as there is no society without gods, there is no society without rumors[1]. Rumor is often defined as the spreading of public interested things, events, or unconfirmed interpretations of problems through various channels[2]. As a way of social communication, rumor has an important function in people's lives, and its most obvious function is to satisfy people's needs for message and emotion[3]. This implies that emotion is also one of the motives behind rumor spreading. Emotion is a kind of psychological activity produced by individuals along with cognition and consciousness, which not only affects an individual's behavior but also can affect the behaviors of other individuals through emotion contagion[4]. For example, In the first week of January 2020, a rumor has linked 5G as being the cause of COVID-19 or accelerating its spread. The issue became a trending topic and appeared visible to all users on Twitter within the UK. Since then, 5G masts were torched in Birmingham and Merseyside due to the emotion of fears that this technology could spread disease [5]. So, it's necessary to study the influence mechanism of emotions on rumor spreading and to avoid inducing destructive behaviors.

[*] corresponding author.
E-mail address:672443312@qq.com(Y.L.L).Full postal address: School of Cultural Management, Nanjing College of Communication, No.3666 Hongjing Road, Nanjing, Jiangsu, 211172, China.

The research on rumor spreading model began in the 1960s, From the macro level, Scholars abstractly classified individuals according to their typical status in the rumor spreading group, and built different models, such as the DK model[6], MK model[7]. Since the 1990s, with the development of information technology and complex networks, the dynamic influence of network topology on rumor spreading has attracted much attention. Zanette D H used the stochastic theory first to establish rumor spreading models on static and dynamic small-world networks respectively[8-9]. Moreno Y et al. set the proportion of different types of populations based on the DK model and proposed the dynamic models of homogeneous and heterogeneous networks rumor spreading in the small-world and scale-free networks respectively[10-11]. However, In contrast to the classic models, Moreno Y et al.introduced the parameter $\alpha$ and $\lambda$ into their model, $\alpha$ means the spreader' interest in or desire for rumors, and $\lambda$ means the forgetting mechanism. Their results showed that not only the network topology has an influence on the rumor spreading, but also the parameter setting has. To be sure, in the real world, no two people have the same experience, in the dissemination of information, each person accepts the story from a slightly different angle and reproduces the story in his way[12]. This determines that the influence of micro variables on rumor spreading cannot be ignored. Thus, Wang et al. studied the influence of the clustering coefficient on rumor spreading [13]. Afassinou K et al. constructed the SEIR model and proposed the inhibitory effect of education rate on rumor spreading[14]. Zhang et al. established the ICSAR model and analyzed eight factors related to rumor spreading[15]. Ma J et al. introduced individual subjective judgment and diversity of individual characteristics to establish a rumor spreading dynamic model in complex heterogeneous networks[16]. Mojgan Askarizadeh et al. took user characteristics, social context, and other factors into account and constructed a rumor evolution game model [17].

Unfortunately, there are still two deficiencies in the existing literature. First of all, emotional variables are not considered, which are regarded as the main factor in the interaction between environmental conditions and human decision-making process [18], and also was proved in public goods games (PGG) [19]. Francoise Reumaux pointed out that rumors intertwine with two lines of connection in his book *Black Widow*. One is the function chain (warning, notification) that is visible and perceived reality. The other is the emotional chain that is invisible, much more complex than the functional chain. Only by using emotions to infiltrate the functional chain can rumors play the suggestive role, so that information can conform to people's expectations [1]. Gustave Le Bon also wrote in *The Crowd* that the instinctive emotions in a group are easily infected with each other and determine the choice of group behavior, which ultimately affects the spreading trend of rumors[20]. Secondly, the literature only studied the conversion mechanism between different population states in rumor spreading but did not take into account the interaction between different involved players. Actually, the essence of rumor spreading is the evolutionary game between different players. Thus, based on the Rank Dependent Expected Utility theory (RDEU), this paper constructs an evolutionary game model of rumor spreading by considering emotion variables, to reveal the influence mechanism of emotions on rumor spreading.

The remainder of this paper is arranged as follows: In section2, an evolutionary game model of rumor spreading considering the emotion variables is constructed. The numerical simulation and correlative analysis are carried out in section 3. At last, the conclusions are presented in section 4.

## 2. A rumor evolutionary game model considering emotional variables

2.1 RDEU theory

RDEU introduced emotional factors into the Expected Utility function, took into account the imperfect rationality of human beings, and overcame the limitation that the traditional Expected Utility Theory cannot fully describe the attitude and degree of economic persons to uncertain risks [21]. Its core is to use "the utility function U $(x)$" and "decision weighting function $\pi(x)$" defines the real-valued function V, to represent the individual preferences of strategy, RDEU model is as follows:

$$V(X, u, \pi) = \sum_{i=1}^{n} \pi(x_i) U(x_i) \qquad (1)$$

Where, for the strategy set X={$x_i$; $i = 1,2,\cdots n$}, the probability of X takes $x_i$ is P {X=$x_i$}=$p_i$, satisfies $p_i \geq 0, p_1+p_2+\cdots+p_n=1$. Sort strategy $x_i$ by U$(x)$ and specify $x_1 > x_2 > \cdots > x_n$. Define the utility level of strategy $x_i$, that is, the Ranking Position (RP$_i$) is:

$$RP_i = P\{X \leq x_i\} = p_i + p_{i+1} + p_{i+2} \ldots + p_n, i = 1,2,3,\ldots n \qquad (2)$$

$$\pi(x_i) = \omega(p_i + 1 - RP_i) - \omega(1 - RP_i), i = 1,2\cdots n[22] \qquad (3)$$

Where, $\omega(\cdot)$ is an emotional function, which can depict the decision-makers emotion [23], and it is a monotonously increasing function satisfying $\omega(0)=0$ and $\omega(1)=1$. The utility theory introduces the hierarchical distribution function and emotion function of strategy, which constitute the cumulative nonlinear decision weight, and can describe the influence of decision-makers' emotional state and intensity under the condition of uncertainty[24].

2.2 Evolutionary game theory

In rumor spreading, the behaviors of decision-makers are affected by some factors and conform to the condition of bounded rationality [25]. They can only adjust their strategies through simple and intuitive decision-making methods such as constant imitation and learning and finally achieve strategic balance. In this paper, a replication dynamic equation model based on the differential equation in evolutionary game theory is applied.

Let the payoff matrix A= $(a_{ij})_n$, and the standard replication dynamic model as follows:

$$\frac{Dp_i}{dt} = p_i[f_i(p) - \bar{f}(p)](1 \leq i \leq n) \qquad (4)$$

Where, $p = (p_1, p_2, \ldots p_n)^T$ is the frequency vector of all strategies, which satisfies $p_1+p_2+\ldots p_n=1$. $p_i$ is the frequency of strategy $i$ in the population, $f_i(p) = \sum_j p_j a_{ij}$ represents the adaptability of strategy $i$ when the system state is $p$, $\bar{f}(p) = \sum_i p_i f_i$ denotes the population average fitness under the state of $p$.

2.3 Model construction and solution

This paper takes netizens(N) and government(G) as decision-makers(DMs), because they are the core relevant players in rumor spreading. Both of them belong to the bounded rational subjects to maximize their interests. In the case of incomplete information symmetry, there are two strategies for netizens, that is spreading(S) and not spreading(NS), and the government's strategies are active monitoring(AM) and negative monitoring(NM). Active monitoring, such as the timely release of authoritative information, proactive response to social concerns, etc. Negative monitoring, including deletion, debate, avoidance, shirking, and other habitual thinking and traditional path dependence.

Assume that the strategic payoff matrix of netizens and government is shown in Table 1, the netizen's payoff satisfies the relation: β>γ=δ>α, and the government's payoff satisfies the relation: ε>θ>η>ζ. Netizens choose to spread rumors with the probability $p(0 \leq p \leq 1)$, and the government takes active monitoring with the probability $q(0 \leq q \leq 1)$. Let $\omega_i(x)=x^{r_i}$, where $r_i$ represents the emotion index of DM $i$, and $r_i > 0$, $i=1,2$, $r_1$ denotes the netizens' emotion index, $r_2$ denotes the government's emotion index. When $0<r_i<1$, $\omega_i(x)$ is a concave function and suggests that the player is optimistic, on the contrary, when $r_i>1$, $\omega_i(x)$ is a convex function, the player holds pessimistic emotion, $r_i = 1$, it indicates that DMs are rational[24]. Thus, the $RP_i$ and $\pi(x_i)$ of netizens and government under each payoff are shown in Table 2.

Table 1. The payoff matrix of DMs.

|   |   | G | |
|---|---|---|---|
|   |   | AM($q$) | NM($1-q$) |
| N | S($p$) | α, ε | β, ζ |
|   | NS($1-p$) | γ, η | δ, θ |

Table 2. The probability distribution, rank position, and decision weight of the strategy payoff values

|   | $x_i$ | $p_i$ | $RP_i$ | $\pi(x_i)$ |
|---|---|---|---|---|
| N | β | $p(1-q)$ | 1 | $[p(1-q)]^{r_1}$ |
|   | γ=δ | $1-p$ | $1-p+pq$ | $(1-pq)^{r_1}-(p-pq)^{r_1}$ |
|   | α | $pq$ | $pq$ | $1-(1-pq)^{r_1}$ |
| G | ε | $pq$ | 1 | $(pq)^{r_2}$ |
|   | θ | $(1-p)(1-q)$ | $1-pq$ | $(1-p-q+2pq)^{r_2}-(pq)^{r_2}$ |
|   | η | $q(1-p)$ | $p+q-2pq$ | $(1-p+pq)^{r_2}-(1-p-q+2pq)^{r_2}$ |
|   | ζ | $p(1-q)$ | $p(1-q)$ | $1-(1-p+pq)^{r_2}$ |

Let U$p$ and E(U$p$) denote the expected utility and average expected utility of netizens adopting S strategy respectively, and U$q$ and E(U$q$) denote the expected utility and average expected utility of governments adopting AM strategy, then we get:

$$U p = \alpha q^{r_2} + \beta(1-q^{r_2}) = \beta + (\alpha-\beta) q^{r_2} \quad (5)$$

$$E(Up)=\beta[p(1-q)]^{r_1}+\gamma[(1-pq)^{r_1}-(p-pq)^{r_1}]+\alpha[1-(1-pq)^{r_1}]$$
$$=(\beta-\gamma)(p-pq)^{r_1} + (\gamma-\alpha)(1-pq)^{r_1}+\alpha \quad (6)$$

$$Uq = \varepsilon p^{r_1} + \eta(1-p^{r_1}) = \eta + (\varepsilon-\eta)p^{r_1} \quad (7)$$

$$E(Uq)=\varepsilon(pq)^{r_2}+\theta[(1-p-q+2pq)^{r_2}-(pq)^{r_2}]+\eta[(1-p+pq)^{r_2}-(1-p-q+2pq)^{r_2}]+\zeta[1-(1-p+pq)^{r_2}]$$
$$= (\varepsilon-\theta)(pq)^{r_2} + (\theta-\eta)(1-p-q+2pq)^{r_2} + (\eta-\zeta)(1-p+pq)^{r_2} +\zeta \quad (8)$$

Thus, we get the replication dynamic equation of rumor spreading by netizens and active monitoring by the government:

$$F(p) = \frac{dp}{dt} = p^{r_1}[Up - E(Up)]$$
$$= p^{r_1}[(\alpha-\beta)(q^{r_2}-1) - (\beta-\gamma)(p-pq)^{r_1} - (\gamma-\alpha)(1-pq)^{r_1}] \quad (9)$$

$$F(q) = \frac{dq}{dt} = q^{r_2}[Uq - E(Uq)]$$
$$= q^{r_2}[(\eta-\zeta)(1-(1-p+pq)^{r_2}) + (\varepsilon-\eta)p^{r_1} - (\varepsilon-\theta)(pq)^{r_2} - (\theta-\eta)(1-p-q+2pq)^{r_2}] \quad (10)$$

So, the replication dynamic model including emotion parameters is

$$\begin{cases} \frac{dp}{dt} = p^{r_1}[(\alpha-\beta)(q^{r_2}-1) - (\beta-\gamma)(p-pq)^{r_1} - (\gamma-\alpha)(1-pq)^{r_1}] \\ \frac{dq}{dt} = q^{r_2}[(\eta-\zeta)(1-(1-p+pq)^{r_2}) + (\varepsilon-\eta)p^{r_1} - (\varepsilon-\theta)(pq)^{r_2} - (\theta-\eta)(1-p-q+2pq)^{r_2}] \end{cases} \quad (11)$$

Let V$p$=U$p$−E(U$p$)=0, V$q$=U$q$−E(U$q$)=0, and the transcendent equations are obtained as follows:

$$\begin{cases} (\alpha-\beta)(q^{r_2}-1) - (\beta-\gamma)(p-pq)^{r_1} - (\gamma-\alpha)(1-pq)^{r_1} = 0 \\ (\eta-\zeta)(1-(1-p+pq)^{r_2}) + (\varepsilon-\eta)p^{r_1} - (\varepsilon-\theta)(pq)^{r_2} - (\theta-\eta)(1-p-q+2pq)^{r_2} = 0 \end{cases} \quad (12)$$

Solve for Eq.(12), and get an equilibrium solution, make it for $(p^*, q^*)$, then the equilibrium solutions of the game are $E_1(0,0), E_2(0,1), E_3(1,0), E_4(1,1)$, and $E_5(p^*, q^*)$. Calculate the derivative of Eq.(11) with respect to $p$ and $q$, and the Jacobian matrix of the replication dynamic model is obtained:

$$J_F(p,q) = \begin{Bmatrix} \frac{dF(p)}{dp} & \frac{dF(p)}{dq} \\ \frac{dF(q)}{dp} & \frac{dF(q)}{dq} \end{Bmatrix}, \text{Tr}(J_F) = (\frac{dF(p)}{dp} + \frac{dF(q)}{dq}), \text{Det}(J_F) = (\frac{dF(p)}{dp} \cdot \frac{dF(q)}{dq} - \frac{dF(p)}{dq} \cdot \frac{dF(q)}{dp})$$

Based on Friedman's method[26], if and only if Det(JF)>0, Tr(JF)<0, the equilibrium point is the evolutionarily stable strategy (ESS) of Eq.(11). In the table below, for simplicity, we make Y indicates that the equilibrium point is an ESS, N indicates it is not, Y/N indicates uncertainty, and SP indicates saddle point.

According to the above analysis, two kinds of situations that are needed to be discussed. First, at least one of the netizens and government is rational, there are three situations: $(r_1=1, r_2\neq 1)$, $(r_1\neq 1, r_2=1)$, $(r_1=1, r_2=1)$. Second, both of them with emotions, the emotional combinations include ($0<r_1<1, 0<r_2<1$), $(r_1>1, r_2>1)$, $(0<r_1<1, r_2>1)$, $(r_1>1, 0<r_2<1)$, which are discussed below.

**Table3.** Evolution situation when $r_1=1$, $r_2\neq 1$.

| Equilibrium Point | $\frac{dF(p)}{dp}$ | $\frac{dF(p)}{dq}$ | $\frac{dF(q)}{dp}$ | $\frac{dF(q)}{dq}$ | Det($J_F$) | Tr($J_F$) | ESS |
|---|---|---|---|---|---|---|---|
| $E_1$ (0,0) | $\beta-\gamma$ | 0 | 0 | 0 | 0 | + | N |
| $E_2$ (0,1) | $\alpha-\gamma$ | 0 | $\varepsilon-\eta$ | 0 | 0 | − | N |
| $E_3$ (1,0) | $\gamma-\beta$ | $\beta-\alpha$ | 0 | 0 | 0 | − | N |
| $E_4$ (1,1) | $\gamma-\alpha$ | $(\beta-\alpha)(1-r_2)$ | $(\varepsilon-\eta)(1-r_2)$ | $r_2(\zeta-\varepsilon)$ | − | +/− | N |
| $E_5$ ($p*(r_2), q*(r_2)$) | It depends on DMs' payoff parameter value and the government's emotions index | | | | | | Y/N |

**Table4.** Evolution situation when $r_1\neq 1$, $r_2=1$.

| Equilibrium Point | $\frac{dF(p)}{dp}$ | $\frac{dF(p)}{dq}$ | $\frac{dF(q)}{dp}$ | $\frac{dF(q)}{dq}$ | Det($J_F$) | Tr ($J_F$) | ESS |
|---|---|---|---|---|---|---|---|
| $E_1$ (0,0) | 0 | 0 | 0 | $\eta-\theta$ | 0 | − | N |
| $E_2$ (0,1) | 0 | 0 | $\eta-\varepsilon$ | $\theta-\eta$ | 0 | + | N |
| $E_3$ (1,0) | $r_1(\gamma-\beta)$ | $(\beta-\alpha)(r_1-1)$ | 0 | $\varepsilon-\zeta$ | − | +/− | N |
| $E_4$ (1,1) | 0 | $\alpha-\beta$ | $(\eta-\varepsilon)(r_1-1)$ | $\zeta-\varepsilon$ | +/− | − | Y/N |
| $E_5$ ($p*(r_1), q*(r_1)$) | It depends on DMs' payoff parameter value and the netizens' emotions index | | | | | | Y/N |

**Table5.** Evolution situation when $r_1=1$, $r_2=1$.

| Equilibrium Point | $\frac{dF(p)}{dp}$ | $\frac{dF(p)}{dq}$ | $\frac{dF(q)}{dp}$ | $\frac{dF(q)}{dq}$ | Det($J_F$) | Tr($J_F$) | ESS |
|---|---|---|---|---|---|---|---|
| $E_1$ (0,0) | $\beta-\gamma$ | 0 | 0 | $\eta-\theta$ | − | +/− | N |
| $E_2$ (0,1) | $\alpha-\gamma$ | 0 | 0 | $\theta-\eta$ | − | +/− | N |
| $E_3$ (1,0) | $\gamma-\beta$ | 0 | 0 | $\varepsilon-\zeta$ | − | +/− | N |
| $E_4$ (1,1) | $\gamma-\alpha$ | 0 | 0 | $\zeta-\varepsilon$ | − | +/− | N |
| $E_5$ ($p^*, q^*$) | 0 | $(\beta-\alpha)(p^2-p)$ | $(\varepsilon-\zeta+\eta-\theta)(q-q^2)$ | 0 | − | 0 | SP |

**Table6.** Evolution situation when $r_1\neq 1$, $r_2\neq 1$.

| Equilibrium Point | $\frac{dF(p)}{dp}$ | $\frac{dF(p)}{dq}$ | $\frac{dF(q)}{dp}$ | $\frac{dF(q)}{dq}$ | Det($J_F$) | Tr($J_F$) | ESS |
|---|---|---|---|---|---|---|---|
| $E_1$ (0,0) | 0 | 0 | 0 | 0 | 0 | 0 | N |
| $E_2$ (0,1) | 0 | 0 | 0 | 0 | 0 | 0 | N |
| $E_3$ (1,0) | $r_1(\gamma-\beta)$ | $r_1(\beta-\alpha)$ | 0 | 0 | 0 | − | N |
| $E_4$ (1,1) | 0 | $r_2(\alpha-\beta)$ | $(\varepsilon-\eta)(r_1-r_2)$ | $r_2(\zeta-\varepsilon)$ | +/− | − | Y/N |
| $E_5(p^*(r_2,q), q^*(r_1,p))$ | It depends on DMs' payoff parameter value and emotions index | | | | | | Y/N |

When $r_1=1$, $r_2\neq 1$, $(p^*, q^*)=(p^*(r_2), q^*(r_2))$, netizens spread rumors with the probability of $p^*(r_2)$, the government choose to actively monitor with the probability of $q^*(r_2)$. The evolution situation of the five equilibrium points is shown in Table 3. $E_1$, $E_2$, $E_3$, and $E_4$ are not ESS, $E_5$ is affected by payoff parameters and the government emotions index, so it is difficult to judge whether it is an ESS, which is discussed in the simulation below.

When $r_1\neq 1$, $r_2=1$, $(p^*, q^*)=(p^*(r_1), q^*(r_1))$. In Table 4, the evolution situation of the five equilibrium points is shown, $E_1$, $E_2$, and $E_3$ are not ESS, E4 satisfies the equilibrium condition and is an ESS when $r_1>1$, but the evolutionary stability cannot be achieved when $r_1<1$, For $E_5$, it is difficult to judge whether it is an ESS because of the unknown DMs' payoff parameter and netizens emotions index.

When $r_1=r_2=1$, namely DMs are all rational, $(p^*, q^*)=(\frac{\theta-\eta}{\varepsilon-\zeta+\theta-\eta}, \frac{\beta-\gamma}{\beta-\alpha})$, netizens choose to spread rumors with $p=\frac{\theta-\eta}{\varepsilon-\zeta+\theta-\eta}$, government choose to actively monitor with $q=\frac{\beta-\gamma}{\beta-\alpha}$. As shown in Table 5, $E_1$, $E_2$, $E_3$, and $E_4$ are not ESS, $E_5$ is a saddle point.

When netizens and the government are in the emotion, that is, $r_1\neq 1$, $r_2\neq 1$, the evolutionary results of $E_1$, $E_2$, $E_3$, $E_4$, and $E_5$ are shown in Table 6. $E_1$, $E_2$, and $E_3$ are not ESS. $E_4$ meets the equilibrium conditions and is an ESS when $r_1>r_2$, otherwise, it's not. In other words, when the netizen's homogeneous sentiment index is larger than the government's homogeneous sentiment index, (S, AM) is a stable solution. Whether $E_5$ is an ESS depends on DMs' payoff parameter value and emotions index.

## 3. Numerical simulation and analysis

According to the analysis of the preceding context, it can be known that the evolutionary stability of the equilibrium point also depends on the payoff parameters and emotional indexes of DMs. Based on the assumption that $\beta>\gamma=\delta>\alpha$, $\varepsilon>\theta>\eta>\zeta$, we refer to Y'method[27] and presuppose that $\alpha=-3$, $\beta=3$, $\gamma=0.5$, $\delta=0.5$, $\varepsilon=1$, $\zeta=-4$, $\eta=-3$, $\theta=0$, then, use MATLAB to simulate and reveal the influence of different emotion types and intensities on the stability of the evolutionary game. Here, the initial value of $p$ and $q$ is 0.5.

3.1. The evolutionary scenario in which at least one decision-maker is emotionally involved.

First, assume that netizens are rational, namely, $r_1=1, r_2\neq 1$. In this case, $r_2$ has two possibilities according to the above. When $0<r_2<1$, it indicates that the government is optimistic, the evolution is shown in Fig.1(a). When $r_2>1$, the government is pessimistic, and the evolution is shown in figure Fig. 1(b).

When the netizens are rational and the government is optimistic, It can be seen from Fig.1(a) that p and q eventually converge to a certain value, forming a stable hybrid strategy. It shows that the more optimistic the government is about the situation, the easier it is to reach internal consensus, the more inclined it is to negative monitoring, the more rational netizens tend to spread rumors to meet their needs, and the closer the evolutionary stability is to a bad combination (S, NM). This shows that if the government is blindly optimistic about the situation, there is a risk that netizens will spread rumors with a high probability and a crisis may erupt. In the long run, it will damage the government's image of social governance, make it lose credibility, and the Tacitus Trap crisis may occur. (Tacitus Trap is named after Tacitus Trap, a historian who lived in ancient Rome. Colloquially speaking, it means that when a government department loses its credibility, whether it speaks the truth or lies, does good or bad, it will be considered to speak lies or do bad things.)

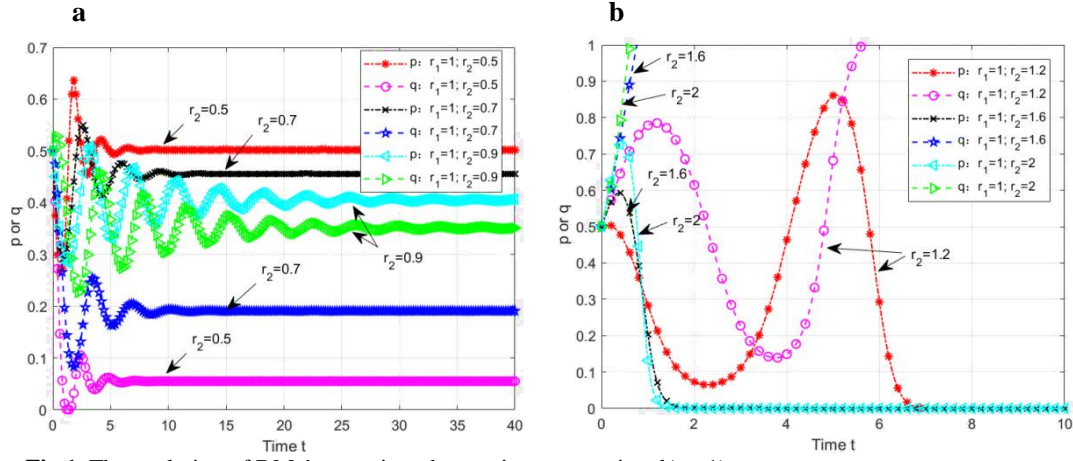

**Fig.1.** The evolution of DMs' strategies when netizens are rational($r_1$=1).

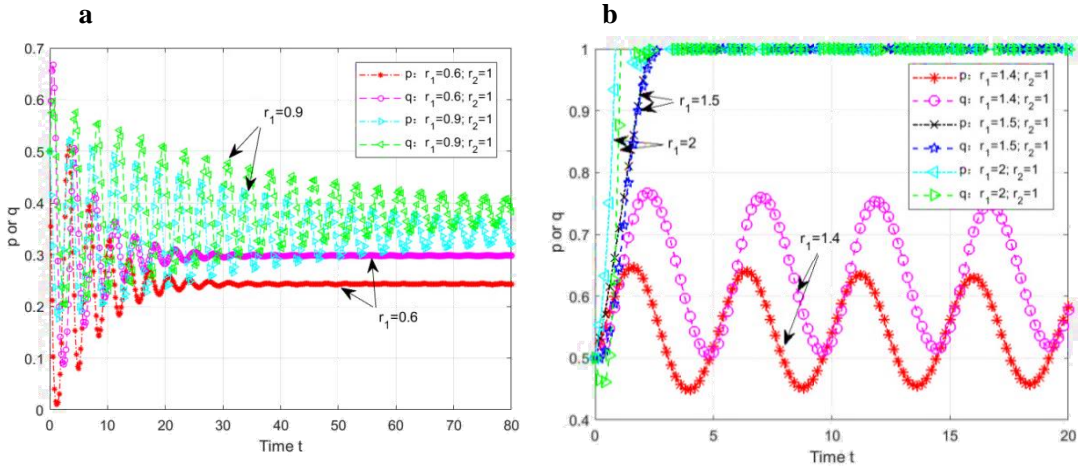

**Fig.2.** The evolution of DMs' strategies when the government is rational($r_2$=1).

As shown in Fig.1(b), when the netizens are rational and the government is pessimistic, pure strategy combination (NS, AM) achieves evolutionary stability and it's an ESS. The stronger the government's emotion is, the faster the convergence speed is. Rational netizens are wait-and-see, adjusting their tactics according to the government's strategy. Although netizens initially spread rumors with a certain probability, they soon stop as the government's active monitoring strategy becomes clear. For example, when $q$ rises to 0.73 ($r_1 = 1, r_2 = 2$), the probability of rumor spreading by netizens drops rapidly from the peak value, and the strategy shifts from S to NS. Through comparison, it is found that if netizens are rational groups, the pessimistic emotion of government is conducive to forming a beneficial situation.

Second, the same as the above, assume that government is rational, that is $r_1 \neq 1$, $r_2 =1$, there are two scenarios for the emotions of netizens. When 0<$r_1$<1, it denotes that the netizens are optimistic, the evolution is shown in Fig.2(a). When $r_1$>1, the netizens are pessimistic, and the evolving situation is shown in Fig.2(b).

A rise in the index of optimism indicates a shift in DM's emotion towards pessimism and vice versa. As shown in Fig.2(a), the more optimistic the netizens are, the easier the strategy evolution is to converge to a stable value. If the optimism index is high, stability cannot be achieved easily. For example, when $r_1$=0.6, the evolution of strategies for S and AM converge rapidly to stable values of 0.24 and 0.29, respectively, while when $r_1$=0.9, evolutionary stability is not formed, and $p<q$<0.5. DMs' strategies change one after another, which reflects that in the interactive decision-making process, they constantly adjust corresponding countermeasures according to the changes in each other's strategies through the learning mechanism. As the emotion rises, the game result tends to form a hybrid strategy (NS, NM) stability, and the rumor can be eliminated by itself in the end.

When $r_1$>1, netizens are pessimistic, as shown in Fig.2(b). In this case, the most important finding is the emotional threshold of 1.4 for netizens to quickly reach a unified decision and spread rumors. When $r_1$≤1.4, the evolutionary game is not stable, and DMs make interactive decisions according to the behavior of the other, and the evolutionary process is similar to Fig.2(a). When $r_1$≥1.5, the evolution rapidly converges to the stability of pure strategy (S, AM), and the netizens reach the decision unity in a shorter time than the government. This shows that when netizens' pessimism is low and still have certain rationality, they are in a wait-and-see state and expect the government to respond positively. However, due to the mechanism barriers, even if the rational government adopts strategy AM with a higher probability than netizens, the information or feeling needs of netizens cannot be satisfied in time. Under the influence of the emotional contagion mechanism, pessimism emotion spreads quickly and causes mass panic and anxiety, netizens quickly reached a unified decision within the group and got involved in the flood of spreading rumors, forming a situation of opposition to the government. In the face of snowflakes of rumors, rational governments rapidly also made unified decisions and adopted strategy AM. But it is unhelpful for the situation due to the netizens are so pessimistic that they have lost all rationality and cannot believe in the government, at last, the evolution result of this case is (S, AM) pure strategy stability, the Tacitus Trap may emerge. Compared with the evolution result brought by netizens' emotions under the premise of rational government, it is found that netizens' optimism is conducive to forming a good situation.

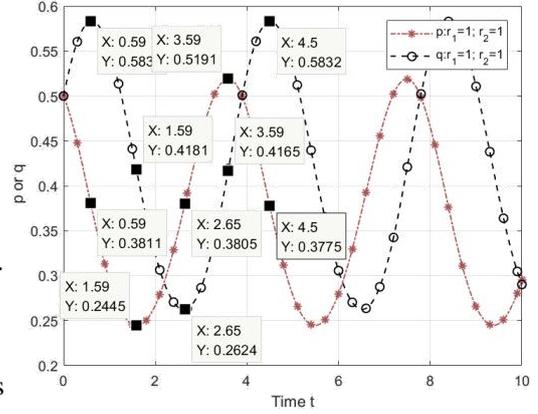

Fig.3.The evolution of DMs are rational($r_1 = r_2$=1)

Third, when DMs hold rational emotions, as shown in Fig.3, the evolutionary game does not converge to a stable value. As time goes by, $p$ and $q$ fluctuate in a fixed amplitude ($p$∈[0.24, 0.51], $q$ ∈ [0.26, 0.58])periodically, which reflects an interactive decision-making process and forms an oscillating situation one after another. $p$=0.38 is the critical value for the government to shift strategy, and $q$=0.41 is the critical value for netizens to change their attitude. For example, $q$ starts to drop from the highest value when $p$ drops to 0.38, and $p$ begins to rebound when $q$ downs to 0.41. And vise versa, when $p$ rises to 0.38, $q$ starts to rebound from the lowest value, when it reaches 0.41, $p$ starts to decline from the highest value. This shows that DMs are influenced by rational emotions when making decisions, act cautiously and do not form clear and unified strategies. This also reflects the learning mechanism of players adjusting corresponding countermeasures according to the transformations of each other's strategies during the game.

3.2.The evolutionary scenario in which DMs are all emotionally involved

As mentioned above, when DMs all have emotions, namely $r_1$≠1, $r_2$≠1, there are emotional combination are (0< $r_1$<1,0<$r_2$<1), ($r_1$>1,$r_2$>1), (0<$r_1$<1,$r_2$>1), ($r_1$>1,0<$r_2$<1), these are discussed below.

The evolution of strategies for DMs with homogenous emotions is shown in Fig. 4. When 0<$r_1$=$r_2$<1, the results are shown in Fig. 4(a), we can get that the evolution converges to a value below 0.5 and $p$ >$q$. With the rise of emotion, the evolution result tends to the strategy combination (NS,

NM), and rumors may go away on their own. If the emotional intensity of one of the DMs is kept constant and changes the other's emotional intensity, some results are shown in Fig4.(b). By comparison with Fig4.(a) and Fig4.(b), it is found that if the emotional intensity of netizens remains unchanged, the more optimistic the government is, the more prone it is to negative monitoring, and the more netizens prone it is to spread rumors. If the emotional intensity of the government remains unchanged, the more optimistic the netizens are, the more they are not to spread rumors, and the more likely the government is to conduct negative surveillance. This further confirms that if the government is blindly optimistic, the probability of netizens spreading rumors will rise, which causes the government to make errors in the analysis and judgment of the situation and miss the opportunity to control rumors, and there may be a hidden danger of Tacitus Trap.

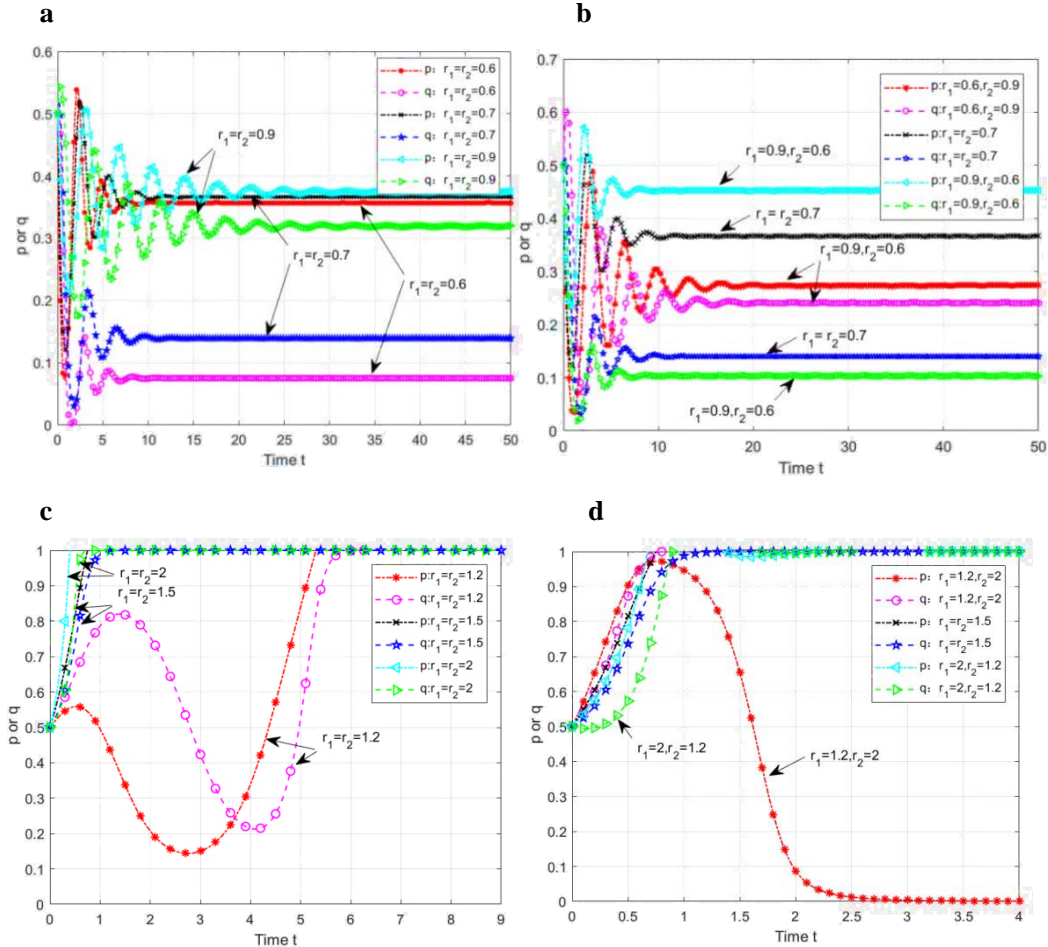

**Fig.4.** Evolutionary scenarios in which DMs hold homogenous emotions. (a)and(b) show the case when the DMs are optimistic, (c)and (d) show the case when the DMs are pessimistic.

When $r_1 = r_2 > 1$, that is DMs with pessimistical emotion, the evolution results are shown in Fig.4 (c), we can get that the evolutionary game of strategies reaches the stability of pure strategic(S, AM) in a very short time, and the larger the emotional index is, the faster it converges. But the more significant finding is that an emotional threshold value of 1.2 at which netizens turn to the opposite strategy due to the different emotional intensities of government. Some results are shown in Fig.4(d) if the emotional intensity of netizens is the same as in Fig.4(c), change the emotional intensity of the government,. By comparing Fig.4(a) and Fig.4(b), if $r_1=r_2=1.2$, strategy evolution converges to the stability of pure strategy(S, AM), netizens choose to spread rumors. However, when $r_1=1.2$, $r_2=2$, $q$ rapidly converges to 1, $p$ decreases from the peak value of 0.97. At this time, the evolutionary game converges to the stability of pure strategy (NS, AM), netizens choose not to spread rumors. This suggests that although netizens are in a pessimistic state, they will have a wait-and-see period on the government's behavior when the sentiment index is low. If the government has a clear and active attitude, they will choose not to spread rumors. This also suggests that the emotion and intensity of government are very sensitive to evolutionary outcomes.

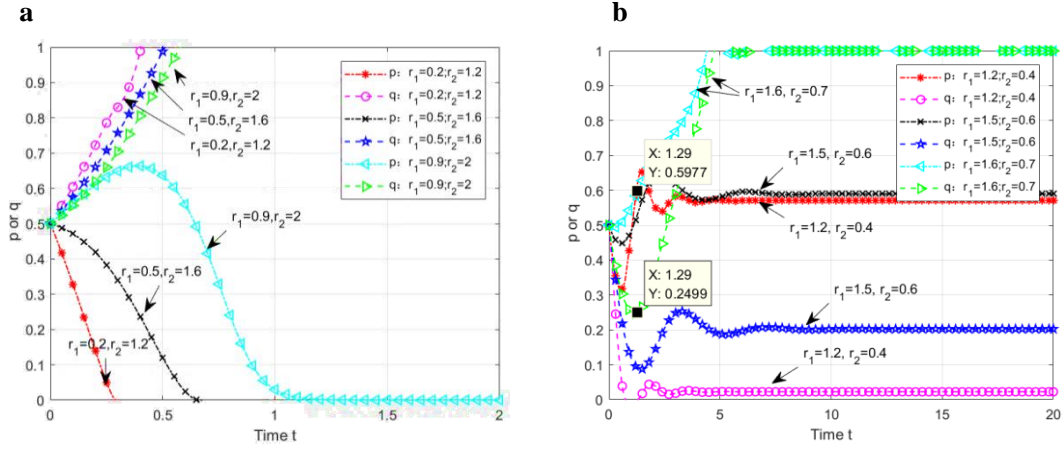

**Fig.5.** An evolutionary scenario in which DMs hold heterogeneous emotions. (a) shows the case when netizens are optimistic and the government is pessimistic, while (b) shows the opposite.

As shown in Fig.5(a), if netizens are optimistic and the government is pessimistic ($0<r_1<1$, $r_2>1$), then the strategy evolution of DMs converges to the stability of pure strategy (NS, AM) in a very short time, which seems to be the most ideal result without any danger due to the preemptive action caused by the government's pessimism.

However, the opposite situation is quite different as shown in Fig5.(b). In this case, one important finding is emotional thresholds, which are 1.6 for netizens and 0.7 for the government. The experiment shows that the evolutionary game converges to the stability of the pure strategy (S, NM) if and only if $r_1 \geq 1.6$ and $r_2 \geq 0.7$. Otherwise, the stability of the hybrid strategy is formed. Besides, another most valuable finding is that $p$ has a threshold value of 0.59 for government strategy conversion under the emotional index combinations ($r_1=1.6, r_2=0.7$), indicating that the government's choice of strategy is not only affected by its own emotions but also affected by whether the probability of netizens spreading rumors exceeds this value. For instance, The $q$ is in a downward trend when $p<0.59$, while it starts to reverse when $p \geq 0.59$, which seems that the optimistic government seems to wake up from the dream and take active monitoring measures rapidly at this time. Finally, the evolutionary game quickly converges to the stability of pure strategy combination (S, AM), and the netizens preceded the government in achieving stability and choose to oppose the government. Altogether, this situation indicates that the negative attitude and blind optimism of the government may make it fall into the passive situation of rumor governance, and the most undesirable situation appears in the evolution result. The government does not avoid falling into Tacitus Trap in the end because it has lost its credibility due to its blind optimism and negative reaction at the beginning of the game.

According to the above analysis, the simulation results of the strategic evolutionary game can be classified into five types: opportunity, opposition, risk, ideal, and security. As shown in Table\ref{tab:my-table7}, if netizens prefer rationality or optimism, the optimistic government will bring risk consequences, while the rational government will bring relatively safe results. If netizens prefer pessimism, no matter what kinds of emotion the government holds, the result may be opportunistic, risky, or oppositional due to the existence of thresholds.

Table 7. The simulation results of strategic evolutionary games

| | | The emotional type of government | | |
|---|---|---|---|---|
| | | optimism ($0<r_2<1$) | pessimism ($r_2>1$) | rationality ($r_2=1$) |
| The emotional type of netizens | rationality ($r_1=1$) | Convergence to hybrid strategy stability, and tend to strategy (S, NM). (risk) | Convergence to the pure strategy (NS, AM) stability. (ideal) | It doesn't converge to stability (security) |
| | optimism ($0<r_1<1$) | Convergence to hybrid strategy stability, and tend to strategy(NS, NM). (risk) | Convergence to the pure strategy(NS, AM) stability. (ideal) | It doesn't converge to stability (security) |
| | pessimism ($r_1>1$) | There are thresholds. When $r_1 \geq 1.8$ and $r_2 \geq 0.6$, Converge to pure strategy (S, AM) stability, otherwise, hybrid strategy stability is formed and tends to (S, NM). (opportunity/risk) | There are thresholds. When $r_1 \leq 1.2$ and $r_2 \geq 2$, Converge to pure strategy (NS, AM) stability, otherwise, converge to pure strategy(S, AM) stability. (opportunity/opposition) | There is a threshold. When $r_1 < 1.5$, Stability is not formed, when $r_1 \geq 1$, convergence to pure strategy (S, NM) stability (opportunity/opposition) |

## 4. Conclusion

Emotions are an indispensable chain of rumors and are considered a major factor in decision-making. This paper takes emotional variables into account and uses the replication dynamic equation to construct a rumor evolutionary game model, and reveal the influence mechanism of type and intensity of emotions on the rumor spreading. Some conclusions are drawn as follows: First of all, it is confirmed that the emotion and intensity preference of DMs have an important influence on the strategy selection in the process of rumor spreading, and different emotional and intensity preferences lead to different strategic choices. The decision-makers' choice of strategy is not only affected by their emotional preference and its intensity but also affected by the other player's. Secondly, the type and intensity of emotion affect the result and stability of strategy evolution, and the emotional combination with a different intensity that DMs prefer will produce different strategy evolution results. Pessimism has a more significant impact on the stability of strategy evolution than optimism, at least one of DMs prefers pessimism, the evolutionary game tends to converge towards pure strategic stability, and its intensity can accelerate the convergence speed. The emotional type of government is more sensitive to the evolutionary stability of strategies. If its emotional preference is optimism, the result of evolutionary strategies tends to be the stability of hybrid strategies, otherwise, it forms the stability of pure strategies. However, the government with rational emotion often chooses and adjusts coping strategies according to the other's emotional strength and strategic choice due to the cautious and conservative decision making, and the evolution of strategies is difficult to form stability. Thirdly, The most interesting and significant finding is some emotional thresholds of strategy change under the pessimistic emotion preference of netizens, which means the window period for netizens to wait-and-see the government's attitude. Meanwhile, it also means an opportunity period for the government to deal with rumors. If the government's nose is sharp enough to get this chance, respond positively and adopt an active monitoring strategy, the \emph{Tacitus Trap} crisis may be defused, or it may fall into it.

**Acknowledgments**

This work is funded by the General Project of Philosophy and Social Science Research in Jiangsu Universities under Grant No. 2020SJA2300.

**References**

[1] Francoise Reumaux，Black widow: the sign and spread of rumors, The Commercial Press, Beijing, 1999.

[2] L.L.Xia, G.P. Jiang, B.Song, Y.R. Song, Rumor spreading model considering hesitating mechanism in complex social networks, Physica A 437 (2015) 295–303.

[3] R.H. Knapp, A Psychology of Rumor, The Public Opinion Quarterly, 8(1) (1944) 22–37.

[4] M Cao, et al., A method of emotion contagion for crowd evacuation, Physica A 483 (2017) 250–258.

[5] W. Ahmed, et al., COVID-19 and the 5G Conspiracy Theory: Social Network Analysis of Twitter Data, J Med Internet Res 22 (5) (2020) e19458.

[6] D.J. Daley, D.G. Kendall, Epidemics and Rumours, Nature 204 (1964) 1118.

[7] D. Maki, M. Thomson, Mathematical models and applications, Prentice-Hall: Englewood Cliff, 1973.

[8] D.H. Zanette, Critical behavior of propagation on small-world networks, Phys. Rev. E, 64 (5) (2001) 050901.

[9] D.H. Zanette, Dynamics of rumor propagation on small-world networks, Phys. Rev. E, 65 (4) (2002) 041908.

[10] Y. Moreno, M. Nekovee, A. F. Pacheco, Dynamics of rumor spreading in complex networks, Phys. Rev. E, 69 (2) (2004) 066130.

[11] Y. Moreno, M. Nekovee, A.Vespignani, Efficiency and reliability of epidemic data dissemination in complex networks, Physical Review E Statal Nonlinear & Soft Matter Physics 69 (5) (2004) 055101.

[12] W. Lippmann, Public Opinion, Wilder Publications, 2010.

[13] Z.F. Pan, X.F. Wang, X. Li, Simulation investigation on rumor spreading on scale-free network with tunable clustering, Journal of System Simulation, 18 (8) (2006) 2346–2348.

[14] K. Afassinou, Analysis of the impact of education rate on the rumor spreading mechanism, Physica A 414 (2014) 43–52.

[15] N. Zhang, H. Huang, et al., Dynamic 8-state ICSAR rumor propagation model considering official rumor refutation, Physica A 415 (1) (2014) 333–346.

[16] J. Ma, H. Zhu, Rumor diffusion in heterogeneous networks by considering the individuals' subjective judgment and diverse characteristics, Physica A 499 (2018) 276–287.

[17] M.B. Askarizadeh, Tork Ladani and M.H. Manshaei, An evolutionary game model for analysis of rumor propagation and control in social networks, Physica A 523 (2019) 21–39.

[18] A. Bechara, A.R. Damasio, The somatic marker hypothesis: A neural theory of economic decision, Games and Economic Behavior 52 (2) (2005) 336–372.

[19] Y.J. Wang. et al., Emotional decisions in structured populations for the evolution of public cooperation, Physica A 468 (2017) 475–481.

[20] L.B. Gustave, The Crowd: A Study of the Popular Mind, Authorhouse, 2008.

[21] J.Quiggin, A theory of anticipated utility, Journal of Economic Behavior and Organization, 3(4) (1982) 323–343.

[22] J. Quiggin, Comparative Statics for Rank-Dependent Expected Utility Theory, Journal of Risk and Uncertainty, 4 (4) (1991) 339–350.

[23] E. Diecidue, P.P. Wakker, On the intuition of rank-dependent utility, Journal of Risk and Uncertainty, 23 (3) (2001) 281–298.

[24] G. Xiong, X. Wang, et al., RDEU Evolutionary Game Model and Simulation of the Network Group Events with Emotional Factors, International Conference on Management Science and Engineering Management. 2018.

[25] D.D. Li, et al., An evolutionary game for the diffusion of rumor in complex networks, Physica A 433 (2015) 51–58.

[26] D Friedman, Evolutionary games in economics, Econometrical, 59 (3) (1991) 637－666.

[27] Y. Yang, J. Wang, Research on the Evolution of Network Public Opinion in Emergencies under the influence of emotional factors, Information science 38 (03) (2020) 35–41,69.